# Improving the reliability of material databases using multiscale approaches

Y. Rollet[a], M. Bonnet[b], N. Carrère[a], F-H. Leroy[a*], J-F. Maire[a*]

[a]Office National d'Etudes et Recherches Aérospatiales, 29 avenue de la division Leclerc, 92322 Châtillon Cedex, France

[b]Laboratoire de Mécanique des Solides, CNRS UMR 7649, Ecole Polytechnique, 91128 Palaiseau Cedex, France

[*]Corresponding authors: fhleroy@onera.fr (F.H. Leroy), Jean-Francois.Maire@onera.fr (J.F. Maire)

**Abstract**

This article addresses the propagation of constitutive uncertainties between scales occurring in the multiscale modelling of fibre-reinforced composites. The amplification of such uncertainties through upward or downward transitions by a homogenisation model is emphasized and exemplified with the Mori-Tanaka model. In particular, the sensitivity to data uncertainty in the inverse determination of constituent parameters based on downward transitions is stressed on an example. Then a database improvement method, which exploits simultaneously the available information on constitutive uncertainties at all scales instead of just propagating those associated with one scale, is presented and shown to yield substantial reductions in uncertainty for both the constitutive parameters and the response of structures. The latter finding is demonstrated on two examples of structures, with significant gains in confidence obtained on both.

## 1. Introduction

Over the past years, various homogenisation methods [1] have been developed in order to determine the macroscopic properties of a heterogeneous material from those of its constituents. Among such materials, polymer matrix composites (PMCs) are a fundamental example of application due to the various observation scales involved (constituents, elementary ply, laminate), the high elastic contrast between their constituents (fibre/matrix), and their intensive and growing use in the aerospace industry.

However, except in few cases [2], multiscale modelling of such materials has up to now been approached only from a purely deterministic viewpoint. A set of material properties at a small scale so determines precisely a corresponding set of material properties at a larger scale. Therefore, most of the time material properties are only known in a certain range because of the inaccuracy of experimental methods and the natural variability of the composite materials induced by imperfectly-controlled manufacturing processes.

The aim of this article is to evaluate the effect of uncertainties at a small scale on the final properties at the larger scales, and to propose a method based on multiscale transitions, referred to here as a database improvement method, for reducing the uncertainties at the various scales. The uncertainty propagation and amplification along scales is stressed in section 2. The proposed database improvement method is then presented in section 3. The confidence gains afforded by the proposed approach for the analysis of composite structures are then demonstrated in section 4 on two examples, namely a hat-shaped component under a three-points bending test, and the buckling of a hole-plate.

2. Propagation of uncertainties and database improvement

Polymer matrix composites can be studied at three different intrinsic scales: (i) the scale of the constituents, fibres and matrix (microscopic scale), (ii) the scale of the unidirectional ply (mesoscopic scale) and finally (iii) the scale of the laminate (macroscopic scale). These three scales are an important characteristic of these materials and underline the usefulness of multiscale methods as predictive tools. For example, the classical laminate theory (CLT) [3] is widely used to estimate the properties of the laminate from those of the unidirectional ply. Likewise, to assess the properties of the ply from those of the constituents, various homogenisation methods (e.g. self-consistent scheme, Mori-Tanaka method…) have been developed. By the way this predictive capacity allows a cost reduction during the design phase of a structure. Indeed, with multiscale methods, the effect of a change in the properties of the fibres or in the properties of the matrix can be determined, and the same for a modification of the stacking sequence. Were the composite material properties perfectly known, none or few experiments would be necessary to redesign a structure. However, this predictive capacity is in practice affected by (i) the difficulty to identify some constitutive properties at the low scale and (ii) the uncertainty range associated to measurements.

*2.1. Case of the micro-to-meso transition (direct -or upward- use)*

The micro-to-meso transition is considered here by using the Mori-Tanaka model [4]. The composite is assumed to consist of an isotropic matrix with properties ($E_m$, $\nu_m$) reinforced by fibres made of a transversely isotropic material with properties ($E_{fl}$, $E_{ft}$, $\nu_{fl}$, $\nu_{ft}$, $G_{fl}$). The fibre volume fraction is denoted $C_f$. By applying a ±10% variation on these properties, uncertainties described on Fig. 1 are obtained at the scale of the lamina. A constrained optimisation method has been used to find the extrema in the admissible interval of the properties of the constituents.

This example emphasizes the amplification by the homogenization method of constitutive uncertainties at a small scale. Uncertainties on the lamina properties appear to be two to three times larger than those on the properties of the elementary constituents. Hence, a mechanism more complex than a simple linear summation of uncertainties is seen to take place.

*2.2. Case of the meso-to-micro transition (inverse -or downward- use)*

Multiscale models are also used in a « downward » way [2,5], for transitions from a larger to a smaller scale. Downward transitions are in particular often used for the determination, from the lamina properties, of fibre properties that cannot be conveniently measured directly. Inverse methods [6] are however known to be very sensitive to uncertainties on the input data. Hence, it is also important to study the amplification of uncertainties by downward transition schemes.

The case of a fibre with an unknown transverse modulus $E_{ft}$ is considered. To identify this property, the value of the transverse modulus $E_{22}^{ply}$ of the lamina, measured at 7500 MPa with an associated uncertainty of ±100 Mpa, is used as input of an inverse method. The remaining constitutive properties are assumed to be perfectly known, and are set to the mean values appearing in Fig. 1.

First, if no uncertainty is considered both on $E_{22}^{ply}$ and the other constituents parameters (Fig. 2a), a unique value is obviously obtained for $E_{ft}$. Next, an uncertainty is considered on $E_{22}^{ply}$ only (Fig. 2b), which does not give rise to major difficulties, the error propagation remaining reasonable for an inverse method. Nevertheless, it can be noted that the uncertainty grows from ±1.3% on $E_{22}^{ply}$ to [-6.9% +7.5%] on $E_{ft}$, and that the mean value of $E_{22}^{ply}$ is not the median value of the uncertainty interval.

Now, if an additional uncertainty on the Poisson's ratio of the matrix ($\nu_m$=0,35±0,02) is considered (Fig. 2c), the error propagation appears to be more significant, with a [-15.7% +20.4%] uncertainty interval for $E_{22}^{ply}$, and becomes really catastrophic (Fig. 2d) if uncertainties on the fibre volume fraction ($C_f$=0.58±0.02) and on the modulus of the matrix ($E_m$=2600±200 MPa) are also taken into account. In the latter case, the uncertainty bounds on $E_{22}^{ply}$ are outside the interval [-35% +200%]. So, small uncertainties on input constitutive properties may lead to very high inaccuracies on constitutive properties identified by means of a downward transition.

3. **Database improvement**

In section 2, the amplification of uncertainties by the use of multiscale models in a direct or inverse way, has clearly been established. This phenomenon appears to cause significant difficulties. However, the existence of available data at all scales and the possibility to make these data interact have not been used in the above-mentioned upward and downward transition schemes. For the upward transition, only the information available at the scale of the constituents were used, while for the downward transition only $E_{22}^{ply}$ was considered at the scale of the lamina.

To minimize the adverse effect of uncertainty amplification through the transition schemes, a database improvement method is now proposed. It is based on a comparison through the multiscale models of all the available data (mean values with uncertainties) at all scales. An improvement on the confidence on the data is expected at all the scales, coming from the elimination of the unsuitable combinations of values of parameters. Indeed, uncertainty intervals on each parameter at the smaller (1) or larger (2) scales define spaces of admissible values (Fig. 3a). These spaces are not perfectly coherent, as emphasized in the uncertainty propagation study. Through multiscale models, the image of the admissible space at scale (1) does not coincide exactly with the admissible space at scale (2) (Fig. 3b). The requirement that parameters should lie within these two admissible spaces leads to define at the smaller scale (Fig. 3c) a new (1)-(2) admissible space (or micro-meso admissible space in the case of the constituents-to-lamina transition). This reduced space is associated with a reduced space (2)-(1) at the larger scale (2).

However the practical implementation of this approach is not so simple. In the case of the micro-to-meso transition through the Mori-Tanaka model, 8 input parameters (the properties of the constituents) have to interact with 5 outputs (the properties of the lamina). Among these data, some are required to lie within narrow intervals. For example this is the case of the fibre volume fraction, or the properties of the lamina (except the $G_{23}$), for which reliable measurements are usually available. But the variability of other parameters is larger, in particular for the $G_{23}$ property of the lamina, considered as inaccessible to direct measurement, as well as for some of the fibre properties (e.g. $E_{ft}$ or $G_{fl}$). Uncertainties on the matrix properties are also present due to their sensitivity to the manufacturing process (cure cycle), but are smaller because of the available experimental data.

Because of the number of parameters (the 8 micro properties) and constraints (intervals prescribed for the 5 lamina properties) considered, optimization methods have proven ineffective. Likewise, defining boundaries in such a complicated eight-dimensional space has proved too difficult. Hence, a method of sampling-selection (Fig. 4) has been used. Points (8-uples) are sampled, regularly or randomly (Monte-Carlo method), in the initial micro space. The points which fit (through the Mori-Tanaka model) with properties of the lamina inside the admissible intervals (determined by measurements) are considered to be admissible.

The material properties of the IM7 / 977-2 are used [7], with micro and meso data given in Table 1 and 2, respectively. The sampling-selection method leads to a substantial reduction in the size (8-dimensional volume) of the admissible space at the micro scale. Indeed, using either the regular and the random sampling approaches, less than 1.3% of the initially chosen sampling points turn out to be meso-admissible.

The set of meso-admissible micro parameters thus obtained has a much smaller 8-dimensional volume than the original set of micro-admissible parameters defined as a Cartesian product of uncertainty intervals. That does not necessarily mean that the uncertainty interval associated to any given micro parameter is reduced by this method. Indeed, in the results shown on Fig. 5, the set of meso-admissible micro parameters is such that the uncertainty range of each individual parameter is almost that initially assumed before the improvement procedure. On the other hand, one also sees that uncertainties affecting different parameters are no longer independent: for a chosen value of one micro property, the respect of constraints at the lamina scale induces a narrower uncertainty range for another micro property. This is how the proposed reduction works. For example, $E_m$ and $G_{fl}$ cannot simultaneously take their respective maximum allowed value.

Fig. 5 might suggest that non meso-admissible micro points lie inside the set of meso-admissible micro parameters. This is because what is actually shown on Fig. 5 is a 2-dimensional projection on the

($E_m$, $G_{fl}$)-plane of the 8-dimensional meso-admissible set. Hence, some of the non meso-admissible 8-uples are such that their ($E_m$, $G_{fl}$)-component is meso-admissible while some other component, e.g. the fibre fraction or the longitudinal fibre modulus, is not. For example, in Table 3, points A and B are 8-uples differing only by the value of their component $E_{ft}$. Hence their projections coincide in Fig. 5, while at the lamina scale $E_{22}$=7280 MPa for point A and $E_{22}$=7550 MPa for point B. Point A is meso-admissible while point B, whose value of $E_{22}$ exceeds the upper bound set at 7400 MPa, is not (Table 2).

## 4. Examples of structural applications

In the previous section, the risks incurred by neglecting uncertainties on the material properties were explained, and the proposed database improvement method has been shown to reduce the lack of knowledge on the micro parameters. In this section, the usefulness of this approach for the analysis of composite structures is now demonstrated.

### 4.1. Three-point bending test of a hat-shaped component

First the case of a hat-shaped structural component subjected to a three-point bending test is considered (Fig. 6). Because of its simplicity, this problem is an appropriate example to demonstrate the effect of the database improvement method in terms of reducing the uncertainty on the response of a structure.

This PMC structure, made of 8 plies with stacking sequence [0°/45°/-45°/90°]$_S$ is submitted to a pressure of 100 MPa applied on its top. The nodes on the two edges are clamped. The variations of the maximum bending value are studied according the values of the material properties. The constitutive data used are again those for the IM7 / 977-2 defined in Table 1, Table 2.

Three sets of material parameters, corresponding to raw (initial non-reduced) micro, raw meso and meso-reduced micro, and identical to those defined in the database improvement method, were used. Each set was taken as input for a batch of computations. Each batch is made of more than 1000 computations, each corresponding to a choice of material properties in the appropriate set. The range of variability for the maximum bending value obtained for each batch of computations is given in Table 4. Fig. 6 shows the deformed profile. These results show that using the reduced set of meso-admissible micro parameters leads to a significant reduction of the uncertainty on the structural response, with a variability estimated at 7.5%, instead of about 26% when the set of raw micro parameters is used.

In addition to showing that a significant improvement in terms of structural response uncertainty is obtained when using meso-admissible micro parameters instead of raw ones, the results also indicate that a lesser but still very substantial improvement is achieved in comparison with results obtained on the basis of lamina properties, the variability estimate in the latter case being 13.7% , i.e. about 80% larger than that based on the reduced micro set. This finding may seem surprising. Indeed, since the meso properties have been used in the database improvement process, one might have expected to find the same variability on the structural response. However, the set of meso admissible properties produced by the database improvement method contains only lamina properties that are achievable from admissible micro properties, i.e. is smaller than the raw meso-admissible set. Hence, all the raw lamina properties are not acceptable, which leads in a reduction of the uncertainty in the structural response.

At this point, it is important to ascertain that the observed uncertainty reduction at the structure level resulting from the use of the reduced set was not caused by insufficiently dense sampling of the micro parameter space. For this purpose, a genetic algorithm has been used to find values of meso-admissible micro properties which maximize or minimize the maximum bending. As it was not realistic to run a FEM computation for each of the 8-uples tested by the genetic algorithm, a response surface approach similar to that of [8-11] has been used. A linear approximation formula linking the maximum bending to the values of lamina properties has been set up, and its coefficients have been adjusted by a least-squares fit to previously-obtained pairs of properties and maximum bending responses. The error associated with the linear approximation formula thus constructed has been found to be acceptable (below 1%). This approximation has been used as the cost function for the genetic algorithm. The bounds on the maximum deflection obtained by this optimization process do not differ significantly from those initially found. Results of FEM simulations performed on the basis of the corresponding material properties agree well the values of the maximum deflection predicted by the linear approximation.

*4.2. Buckling of a hole-plate*

Since the proposed approach and the various associated methods have proved beneficial when applied to the hat-shaped component, a more complex case is now considered, namely the buckling of a hole-plate, where geometric non-linear effects are therefore present. The mesh of the component and the boundary conditions are depicted in Fig. 7. Layered finite elements are used. The material properties used are again those of the IM7/977-2, the stacking sequence being $[45°/0°/90°/-45°]_S$.

The computational cost of this structural analysis does not allow to carry out thousands of such FEM computations. Hence, 32 sets of meso-admissible micro properties have been selected such that their corresponding values at the lamina scale are close to the admissibility bounds in the meso space. Three other sets of properties, chosen close to the centre of the admissible set, are added. In all, 35 FEM analyses have been performed. Based on the conclusion of the previous case, this number is a good compromise between the computational cost and the quality of the approximation. The deformed shape obtained is presented Fig. 7.

The force-displacement response curve is now considered. This curve can be described by a four-parameter model ( $f = a \cdot u + b \cdot \langle u - u_0 \rangle^n$, where *a* is the slope of the elastic domain, $u_0$ the buckling threshold and $\langle z \rangle = (z+|z|)/2$ is the positive part (Macauley bracket). Following the method used in the case of the hat-shaped component, a linear approximation is defined to link *a, b, n, $u_0$* to the lamina properties. The approximation is identified on the 35 computations. Since the estimated error is reasonable, the obtained approximation formula is used to simulate the buckling response of the structures for about 1000 samples of material properties chosen in each of the sets of raw micro, raw meso and reduced parameters. Fig. 8-10 and Table 5 have been obtained.

On this example, the use of the database improvement method again yields substantial benefits. The envelope of the curves obtained on the basis of the raw-micro admissible set (Fig. 8) is wider than those for the two other admissible sets. The lower limit of the envelope for the raw-meso admissible set (Fig. 9) coincides with that of the raw-micro envelope. The micro-meso reduced space data span a thinner envelope whose upper limit coincides with that of the raw-meso envelope (Fig. 10). It appears from the results of Table 5 that a single indicator such as the buckling force does not necessarily reflect very well the reduction. In particular, the difference in uncertainties on the buckling force induced by the raw-micro and raw-meso admissible parameter sets is small, whereas Fig. 9 shows an overall significant reduction in response uncertainty when the latter set is used.

## 5. Conclusions

This article has stressed the importance of taking into account constitutive uncertainties in multiscale modelling of composite materials. Such uncertainties undergo considerable amplification as they are propagated upwards or downwards, and cannot be accurately predicted by a simple linear summation. More elaborate schemes are thus necessary for evaluating and reducing the uncertainties at the various scales of the composite material.

The proposed database improvement method, which exploits simultaneously the available informations on constitutive uncertainties at all scales instead of just propagating those associated with the smallest or largest scale, leads to reduced and more realistic estimations of uncertainties at both the micro and meso scales of the composite. Applied in association with FEM structural analyses, the proposed approach is shown to lead to reduced uncertainties on the structural response as well. The proposed approach thus constitutes a possible rational basis for reducing security coefficients used in the design phase and thus avoiding the oversizing of composite structures.

The proposed approach so far accounts only for variability of material properties which are spatially uniform. The next step consists of developing similar methods for the assessment of the effect of spatially non-uniform variabilities (variations in space of a given parameter), which is also an important issue for polymer matrix composites.

**Acknowledgments**

This work has been carried out within the AMERICO project (Multiscale Analyses: Innovating Research for Composites) led by ONERA. Funding by the DGA/STTC (French Ministry of Defence) is gratefully acknowledged.

Micro scale uncertainties    Meso scale uncertainties

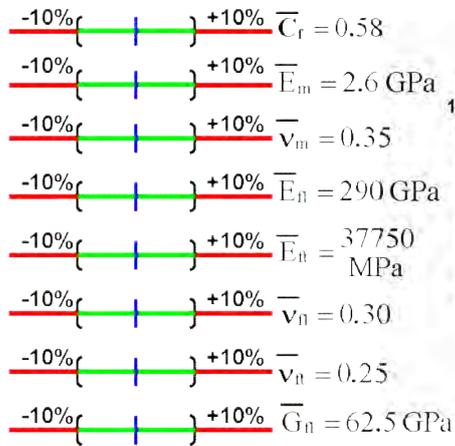
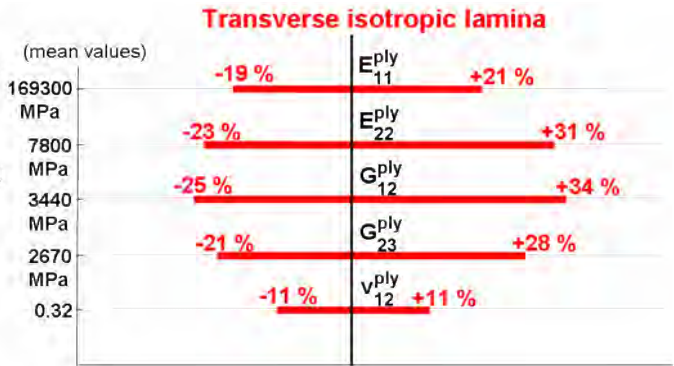

Fig. 1. Growth of material uncertainties from micro to meso scale.

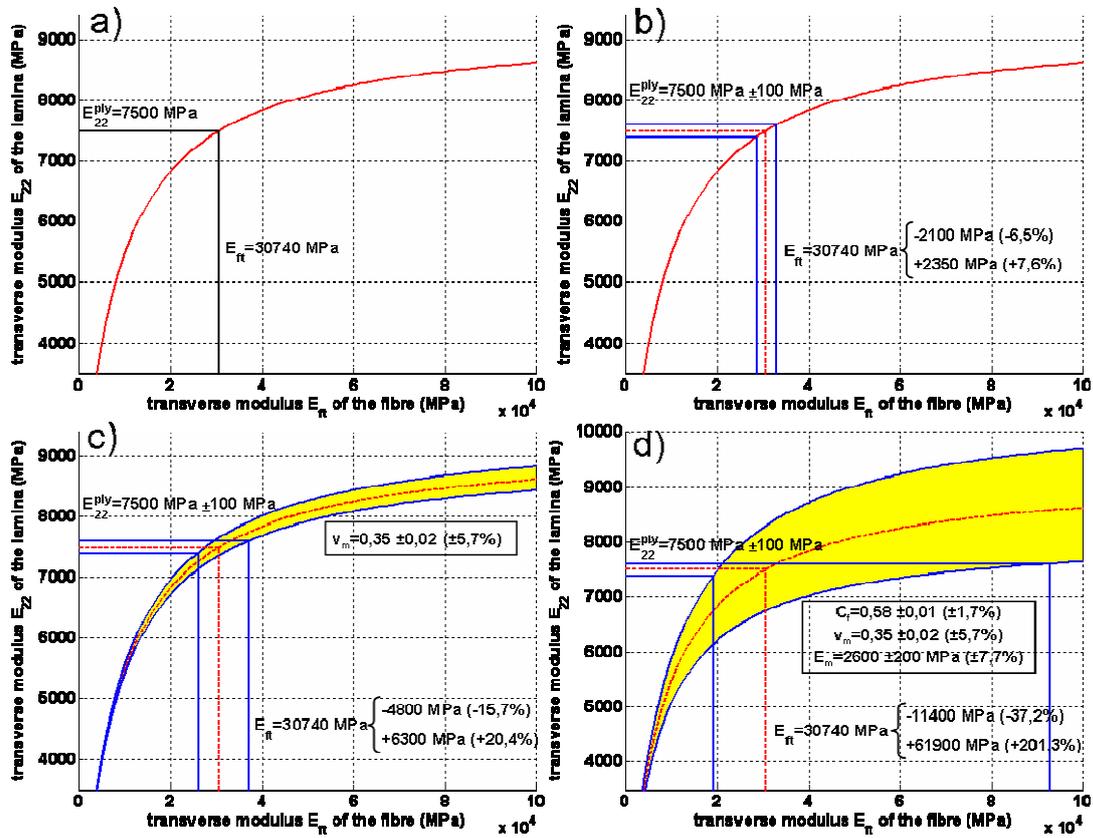

Fig. 2. Effects of material uncertainties in the inverse determination of the transverse modulus of the fibres $E_{ft}$. In a first step, the $E_{22}^{ply}$ and all the micro properties are known (a) whereas in the next step uncertainties are considered on (b) $E_{22}^{ply}$, on (c) $E_{22}^{ply}$ and $\nu_m$ and finally on (d) $E_{22}^{ply}$, $C_f$, $\nu_m$ and $E_m$.

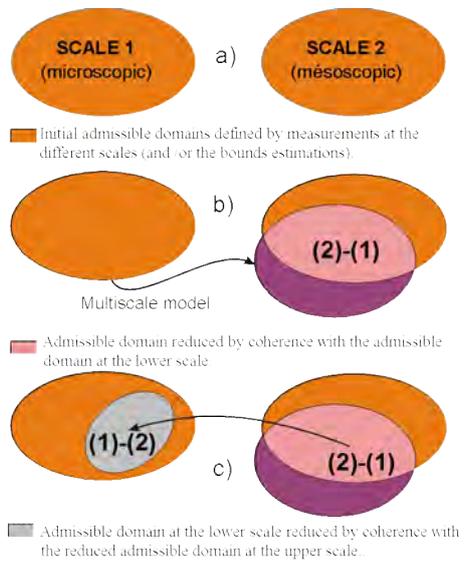

Fig. 3. Principle of databases improvement.

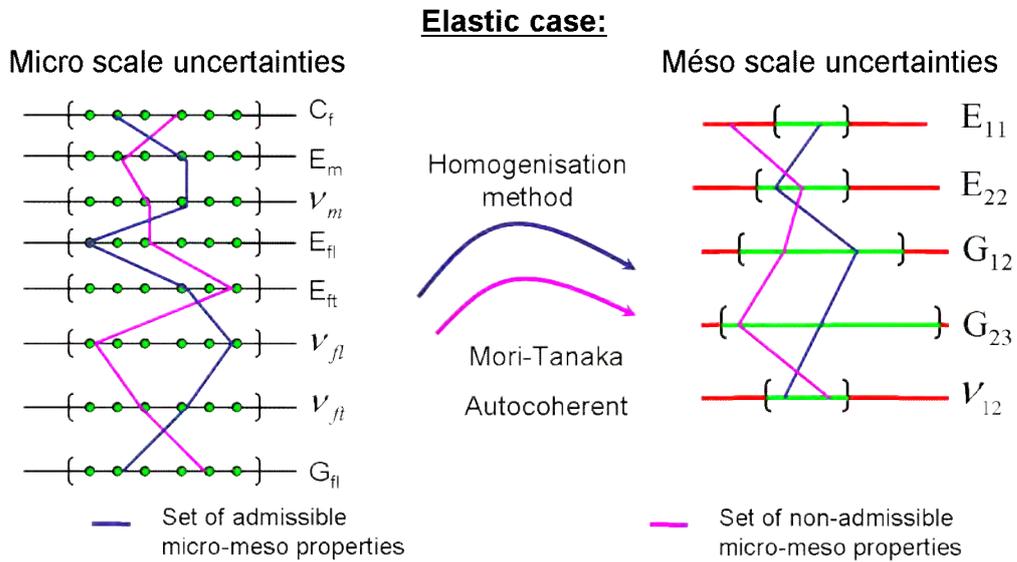

Fig. 4. Principle of sampling-selection.

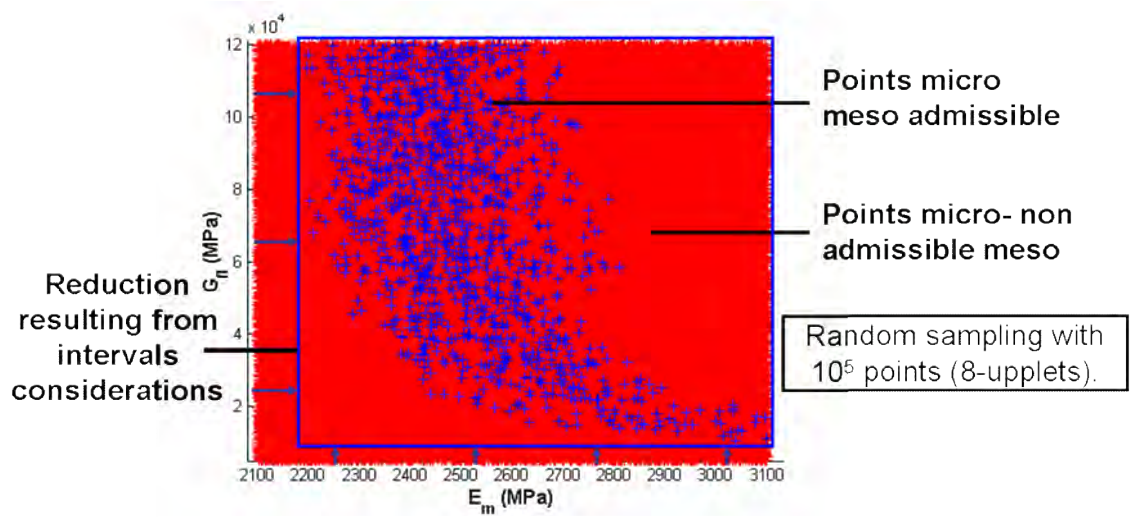

Fig. 5. Micro-meso reduced space for the couple of parameters ($E_m$, $G_{fI}$).

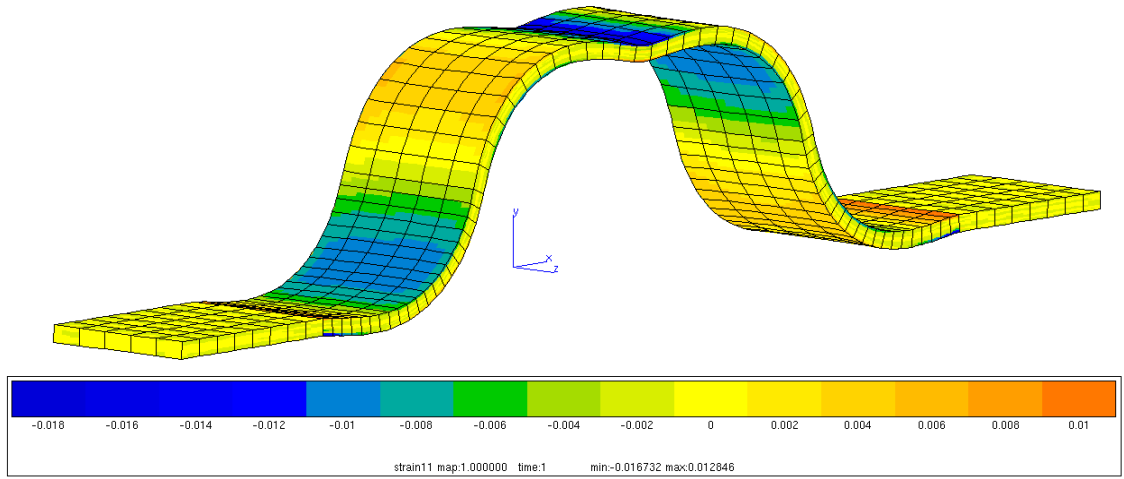

Fig. 6. $\varepsilon_{11}$ field in the ply axes for the deformed "hat-shape" component (displacement x3).

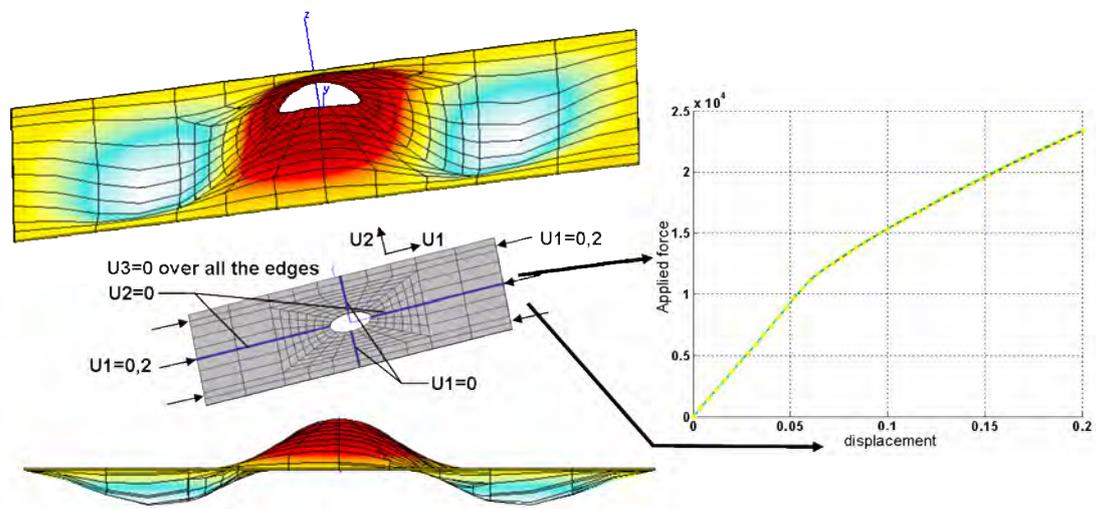

Fig. 7. Deformed shape and response curve for the hole plate under buckling.

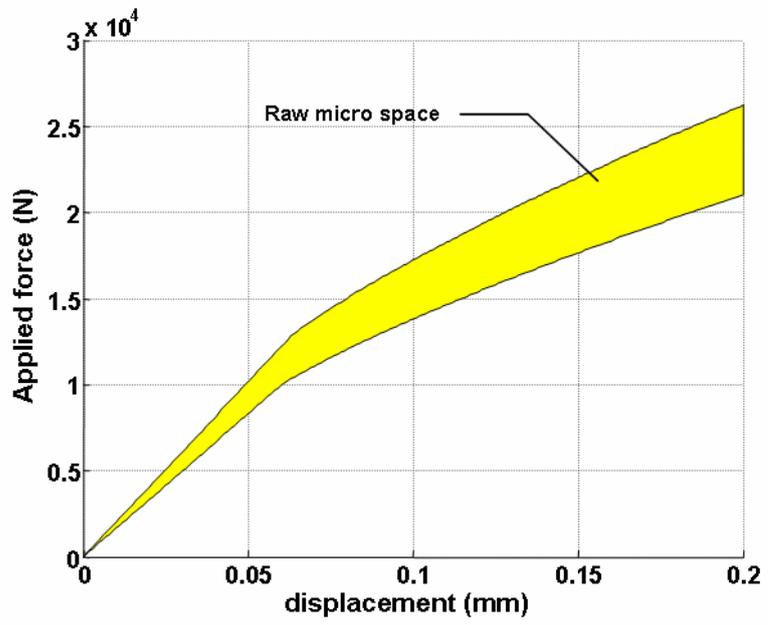

Fig. 8. Variability of the plate response induced by uncertainty on the micro properties (raw micro set).

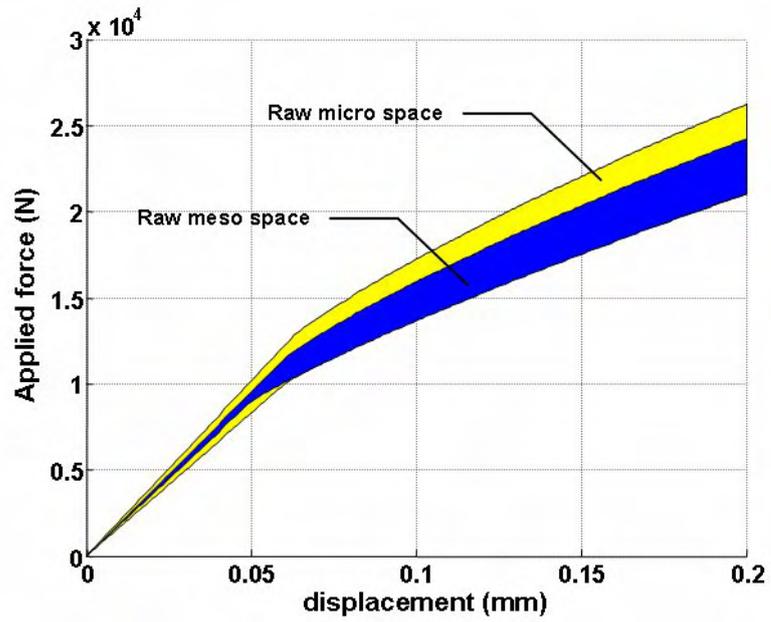

Fig. 9. Comparison of ranges of variability of the plate response induced by uncertainties on the micro

properties (raw micro set) or on the meso properties (raw meso set).

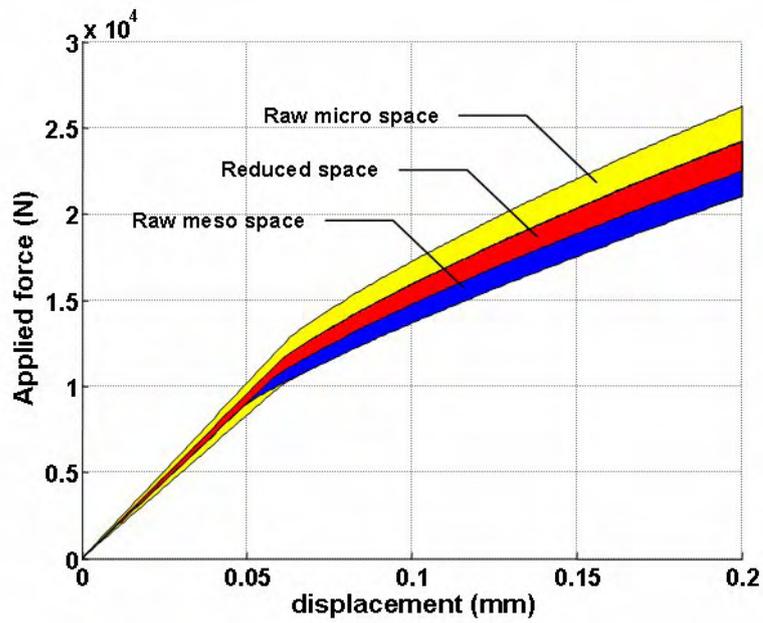

Fig. 10. Comparison of ranges of variability of the plate response induced by uncertainties on the micro properties (raw micro set), the meso properties (raw meso set) or the meso-admissible micro properties (reduced set).

|  |  | Minimum | Maximum |
|---|---|---|---|
| Matrix (isotropic) | $C_f$ | 0.56 | 0.6 |
|  | $E_m$ (MPa) | 2100 | 3100 |
|  | $\nu_m$ | 0.3 | 0.45 |
| Fibre (transverse isotropic) | $E_{fl}$ (MPa) | 270000 | 310000 |
|  | $E_{ft}$ (MPa) | 13500 | 62000 |
|  | $\nu_{fl}$ | 0.23 | 0.33 |
|  | $\nu_{ft}$ | 0 | 0.5 |
|  | $G_{fl}$ (MPa) | 5000 | 120000 |

Table 1 - Uncertainty intervals on the constituent properties.

|  | Minimum | Maximum |
|---|---|---|
| $E_{11}$ (MPa) | 165 000 | 175 000 |
| $E_{22}$ (MPa) | 6600 | 7400 |
| $\nu_{12}$ | 0.28 | 0.34 |
| $G_{12}$ (MPa) | 3200 | 3400 |
| $G_{23}$ (MPa) | 2000 | 4000 |

Table 2- Uncertainty intervals on the lamina properties.

| micro scale properties | $C_f$ | $E_m$ | $\nu_m$ | $E_{fl}$ | $E_{ft}$ | $\nu_{fl}$ | $\nu_{fl}$ | $G_{fl}$ |
|---|---|---|---|---|---|---|---|---|
| Point A | 0.58 | 2500 MPa | 0.33 | 290 GPa | 35 GPa | 0.27 | 0.40 | 65000 MPa |
| Point B | | | | | 45 GPa | | | |

| meso scale corresponding properties | $E_{11}$ (MPa) | $E_{22}$ (MPa) | $\nu_{12}$ | $G_{12}$ (MPa) | $G_{23}$ (MPa) |
|---|---|---|---|---|---|
| Point A | 169250 | 7280 | 0.29 | 3336 | 2522 |
| Point B | | *7550>7400* | | | 2615 |

Table 3 - Example of admissible (*A*) and non-admissible (*B*) points which coincide in the ($E_m$, $G_{fl}$)-plane.

|  | Raw micro | Raw méso | Reduced |
|---|---|---|---|
| Maximum bending | -2,3694 mm | -2,0987 mm | -2,0843 mm |
| Minimum bending | -1,7455 mm | -1,8105 mm | -1,9271 mm |
| Relative variability | **26,00%** | **13,73%** | **7,50%** |
| Absolute variability | 0,6239 mm | 0,2882 mm | 0,1572 mm |

Table 4 - Variability of maximum deflection according to material parameters admissible sets used.

|  | Raw micro | Raw méso | Reduced |
| --- | --- | --- | --- |
| Maximum buckling force $M$ | 12687 N | 11597 N | 11542 N |
| Minimum buckling force $m$ | 9947 N | 8866 N | 10450 N |
| Uncertainty $\varepsilon$ | **24,3%** | **24,2%** | **9,7%** |

With $\varepsilon = 100 \times \dfrac{M - m}{f_{moy}}$ and $f_{moy}$=11276 N

Table 5 – Uncertainties on the buckling force according to material parameters admissible sets used.